# Exploring properties of the integrating pixels

P.J.Kapusta[a], Sz. Bugiel[b] ,R.Dasgupta[b] , S.Glab[b] , M.Idzik[b] , W.Kucewicz[b], M.Turala[a]

Y.Arai[c] , T.Miyoshi[c]

[a]*Institute of Nuclear Physics PAN, Krakow, Poland*
[b]*AGH University of Science and Technology, Krakow, Poland*
[c]*High Energy Accelerator Research Organization(KEK), Tsukuba, Japan*

## 1. Introduction

This document presents some observations and thoughts, which appeared during tests of the SOI sensors based on the integration type pixels.
It contains two parts:

- rough analysis of the *Correlated Double Sampling* filtering properties with respect to different noise sources and long sampling intervals, typical for the pixels under consideration;
- results of the pixel leakage current measurements in the *pix_2012* and *DIPIX* pixel detector chips.

## 2. Integration type pixels

Schematic diagrams of the pixels used in the *DIPIX* [2] chips and also in the *pix_2012* [3] chip from Krakow are shown in Drawing 1. Signal charge is integrated on the input capacitance during defined, constant periods lasting up to few hundreds of us. Each integration period is followed by a short reset, necessary to remove the collected charge. A N-type source follower provides current amplification of the signal. Two capacitors, each about 100fF are used for storing signal samples at the start and end of the integration period.

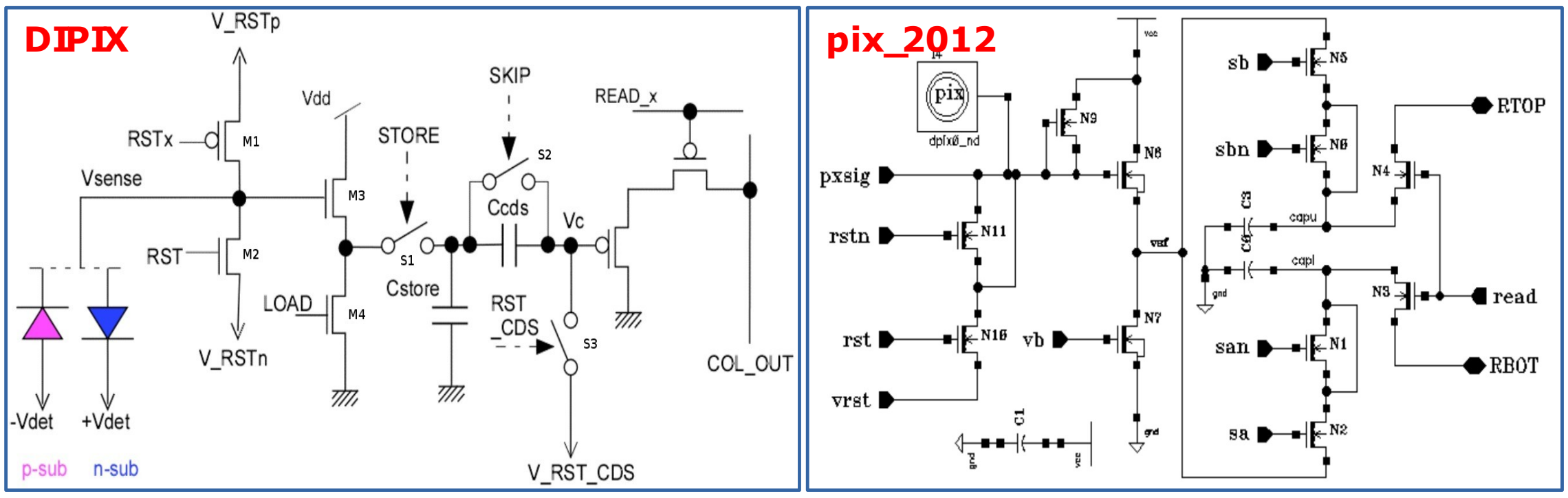

*Drawing 1: Schematic diagrams of the pixels in the DIPIX and pix_2012 chips.*

Pixel operation is driven by three switches:

- RESET (RST, rst), connected to pixel capacitance;
- First Sample (RST_CDS, sa);
- Second Sample (STORE, sb);

The First Sample is subtracted from the Second one, what is usually called as the CDS – *Correlated Double Sampling*. Note, that the source follower load is changed between sampling operations – the first sample is taken with the load of 200fF, while the second one with the 100fF only.

## 3. Filtering properties of the *Correlated Double Sampling*

### 3.1 CDS and the parallel noise

The pixel architecture presented above may be considered as an example of the *time-variant* filter, frequently presented in literature [1]. The idea is shown in Drawing 2; Drawing 3 contains a timing diagram adequate for the noise calculations.

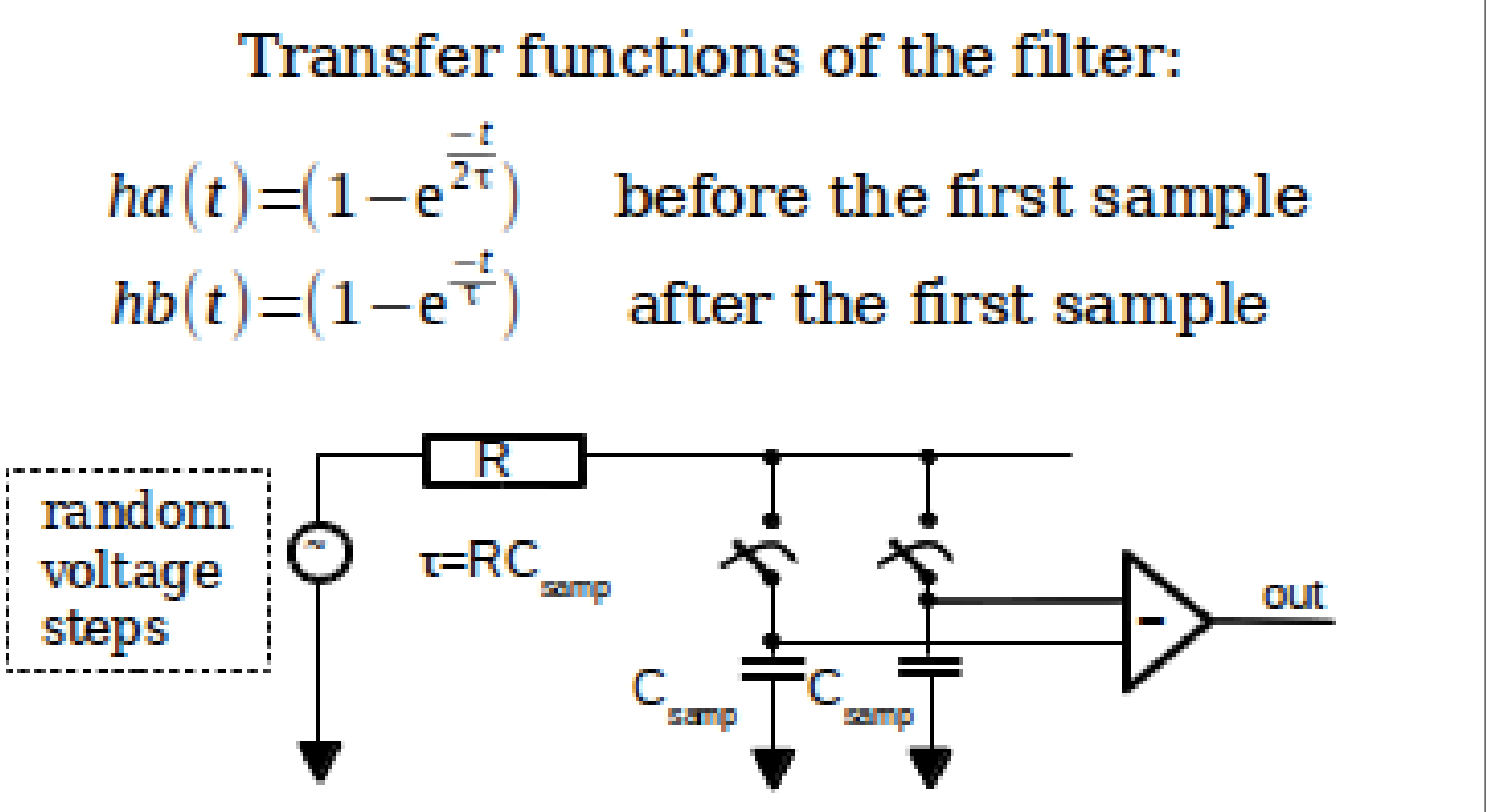

*Drawing 2: Time-variant CDS filtering of the parallel noise.*

The noise source in Drawing 2 produces voltage steps, which result from the current “Dirac” pulses arriving on the pixel capacitance.

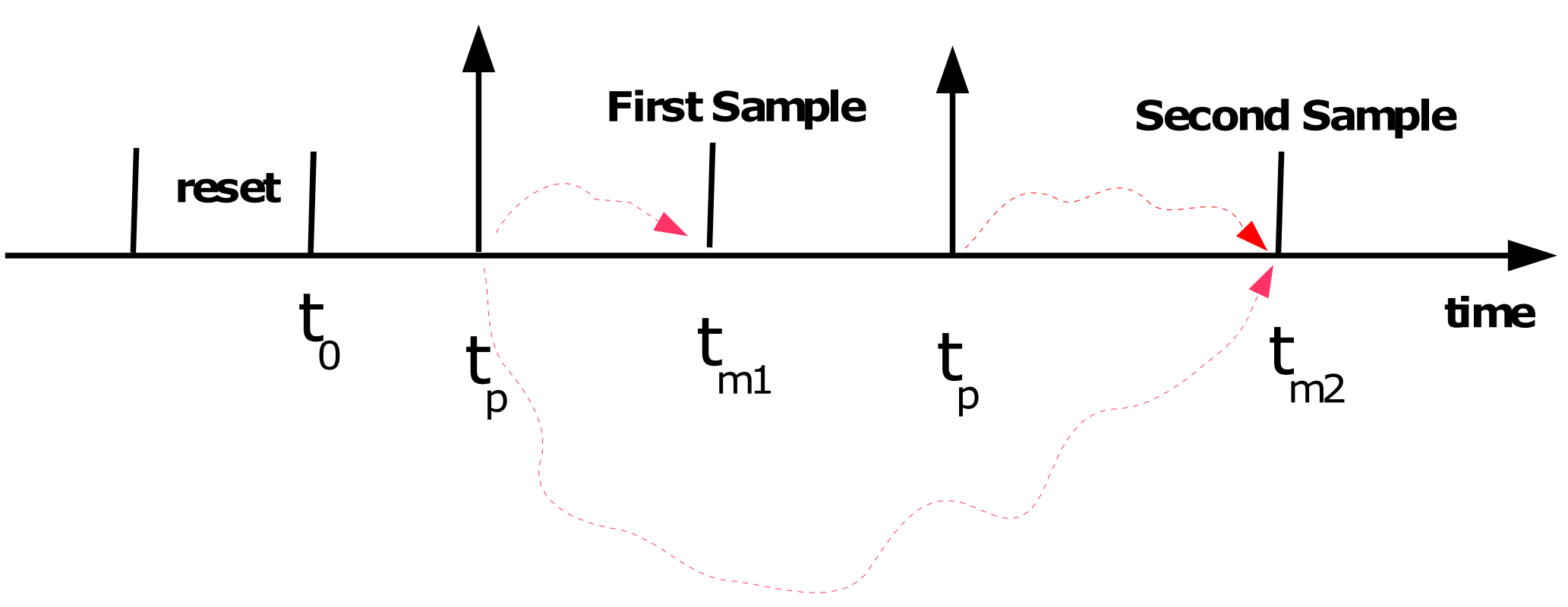

*Drawing 3: Timing diagram relevant to the noise calculation.*

We calculate a total contribution from noise pulses produced at times tp. Pulses between t0 and tm1 affect both tm1 and tm2 samples, while pulses between t1 and tm2 affect only tm2 sample.
The time constant τ (Drawing 2) depends on the sampling capacitance, transconductance of the input transistor and the resistance of the sampling switch:

$$\tau = RC = C_{samp}\left(\frac{1}{g_m} + R_{sw}\right)$$

To find the *Equivalent Noise Charge* (*ENC*) we integrate squared noise contributions in the interval from $t_0$ to $t_{m2}$ using transfer functions hA(t) and hB(t), based on the ha(t) and hb(t) ones:

$$ENC^2 = A\left(\int_0^{tm1} hA(t)^2 dt + \int_{tm1}^{tm2} hB(t)^2 dt\right)$$

where the “A” factor is an unilateral parallel noise density.

$hA(t_p) = (1 - ha(tm_1 - t_p))\, hb(tm_2 - tm_1)$ , for $t_p$ pulses between 0 and $t_{m1}$;

$hB(t_p)=hb(tm_2-t_p)$ , for $t_p$ pulses between $t_{m1}$ and $t_{m2}$

Given above formulas lead to:

$$ENC=\left(i_{det}\cdot e\cdot\left(tm2-tm1-\frac{3\tau}{2}+2\tau e^{\frac{tm1-tm2}{\tau}}-\frac{1}{2}\tau e^{\frac{2(tm1-tm2)}{\tau}}+\left(-1+e^{\frac{tm1-tm2}{\tau}}\right)^2\left(\tau-e^{\frac{-tm1}{\tau}}\tau\right)\right)\right)^{0.5}$$

The interval $T=tm2-tm1$ is usually called the „integration time". Since the $T >> \tau$ (integration time is in range of hundreds microseconds), the *ENC* formula can be simplified to:

$$ENC=(i_{det}\cdot e\cdot T)^{0.5}$$

where $i_{det}$ is a current from which the parallel noise originates and the e represents an electron charge.

### 3.2 CDS and the White Noise

The exact calculations of the *time-variant* filters in time-domain are not practical in the engineering practice. Due to this the frequency-domain approach will be used in the following material, while keeping in mind that obtained result are approximate.

In the frequency-domain the CDS filtering is represented as a transfer function:

$$|H(\omega)|^2=2-2\cos(T\omega)$$

We will consider now the first order low-pass filter of the time constant τ. The "whole bandwidth" noise at the output:

$$u^2_{nocds}=WF\frac{1}{2\pi}\int_0^\infty\frac{1}{1+w^2\tau^2}d\omega=WF\frac{1}{4\tau}$$

where WF is a constant factor, depending on the noise origin.

After adding the CDS stage we get:

$$u_{cds}^2 = WF \frac{1}{2\pi} \int_0^{\infty} \frac{1}{1+w^2\tau^2} \left(2-2\cos(T\omega)\right) d\omega = WF \frac{1-e^{\frac{-T}{\tau}}}{2\tau}$$

A ratio :

$$\frac{u_{cds}^2}{u_{nocds}^2} = 2-2e^{\frac{-T}{\tau}} \sim 2$$

Since the $T >> \tau$, the CDS stage practically doubles the white noise in the system.

### 3.3 CDS and the Flicker Noise

Calculations involving the flicker noise are more difficult then the ones already presented. Drawing 4 can help in understanding how the CDS process the flicker noise. Here the squared modulus of the CDS transfer function relevant to pixel

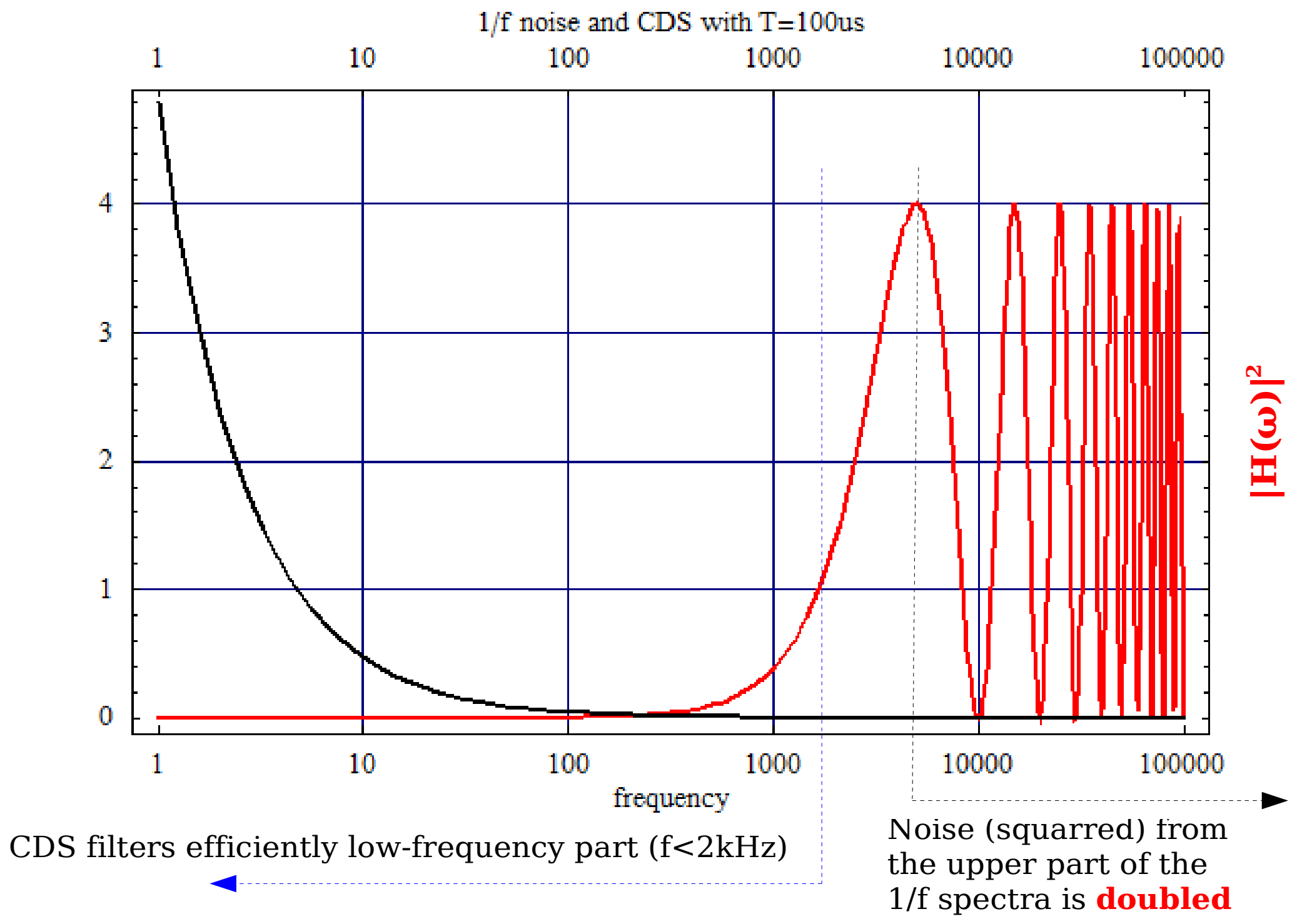

*Drawing 4: Qualitative illustration of the CDS filtering properties with repect to the flicker noise.*

operation (integration time T=100us) is superimposed on the *1/f* noise spectra. The CDS efficiently filters all noise components below 2kHz (it is a principal advantage of this filtering). However, the upper part of the spectra is partially attenuated and partially amplified. The squared *1/f* noise, integrated in a single decade, is equal $ln(10)/2\pi$ . With the CDS filtering applied this value can be found as $ln(10)/\pi$ for high frequencies, i.e. the CDS **doubles** the squared *1/f* noise for $f >> 1/T$ .
Let's now consider the first order low-pass filter followed by the CDS filtering stage. The output noise of such a system can be found as:

$$u^2=\frac{1}{2\pi}\int_0^{\infty}\frac{1}{1+\omega^2\tau^2}\frac{2-2\cos(T\omega)}{\omega}d\omega$$

Above integral leads to expression:

$$u^2=\frac{1}{2\pi}\left(2\mathrm{EulerGamma}+\ln\left(\frac{T^2}{\tau^2}\right)-2\cosh\left(\frac{T}{\tau}\right)coshIntegral\left(\frac{T}{\tau}\right)+2\sinh\left(\frac{T}{\tau}\right)sinhIntegral\left(\frac{T}{\tau}\right)\right)$$

For the $T/\tau >10$, only two first elements of the sum are important and noise can be approximated:

$$u^2=\frac{1}{2\pi}\left(2\mathrm{EulerGamma}+\ln\left(\frac{T^2}{\tau^2}\right)\right)$$

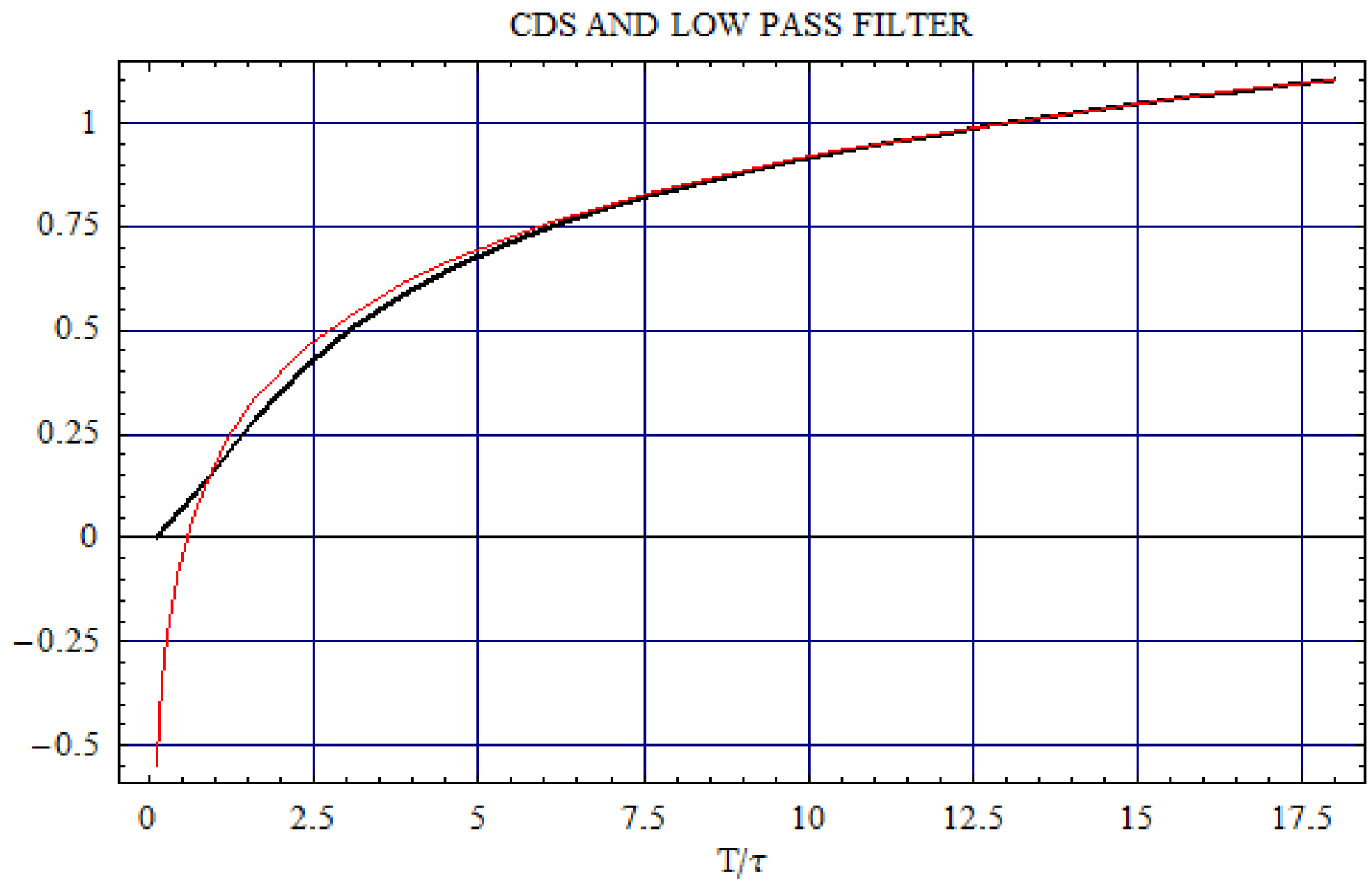

*Drawing 5: Output noise of the system composed of the 1/f noise source and the low-pass filter appended with the CDS stage. T - integration time, τ — time constant of the filter. Black curve — the exact calculation, red one — the approximation.*

Corresponding curves are shown in Drawing 5. Longer integration times correspond to a higher level of the passing *1/f* noise. The noise slope can be found as : $1/\pi T$.

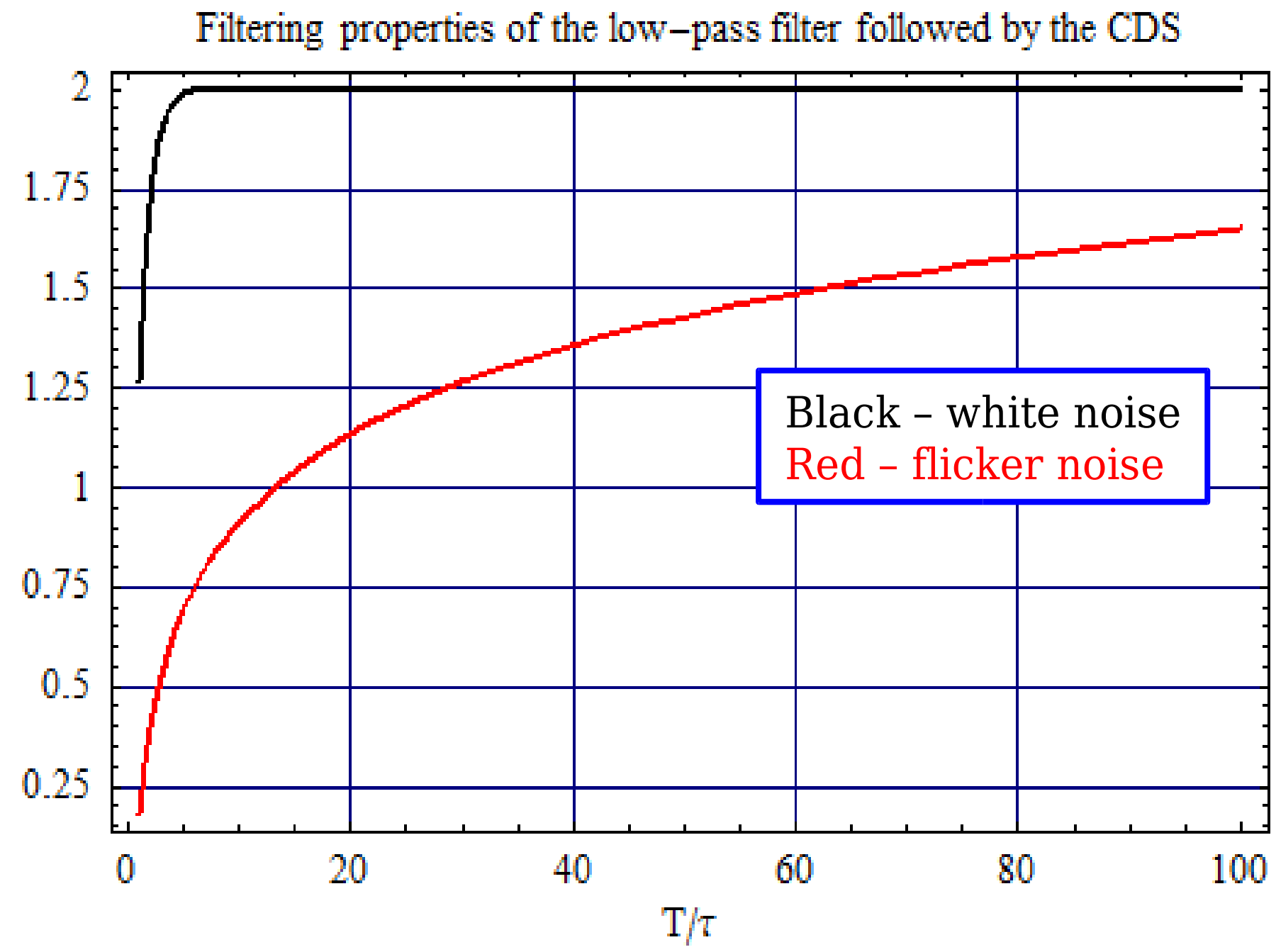

*Drawing 6: CDS applied to the white and flicker noise.*

It is instructive to compare on a single picture output noise of the low-pass filter appended with the CDS sourced with the white and flicker noise, as in the Drawing 6. The illustration is purely qualitative – the white noise curve corresponds to the ratio (see section 3.2) while the flicker noise curve is a squared voltage obtained with spectral density equal unity. However, it is important to note that only for very short „integration times" (hundreds of nanoseconds) the CDS attenuates the white noise. On the other hand, the flicker noise manifests itself in the visible, monotonous noise versus integration time dependency.

Unfortunately, the relevant calculations for *1/f* noise with the flicker noise exponent *AF* not equal 1 :

$$u^2 = \frac{1}{2\pi} \int_0^{\infty} \frac{1}{1+\omega^2 \tau^2} \frac{2-2\cos(T\omega)}{\omega^{AF}} d\omega$$

are not simple and have to be made numerically.

## 3.4 Decomposition of the *DIPIX* FZp floor noise

The floor („electronic”) noise of the pixel front-end alone (the Source Follower and the CDS) was measured for different integration times in range from tens of nanosecond up to 4ms. The pixel was kept at permanent reset. The readout noise (chip and external amplifiers, ADC) was subtracted (measurement of this noise is possible with the RST_CDS switch permanently “on”, see Drawing 1). Resulted curve was fitted to a sum of the white and flicker noise components, accordingly to given previously formulas.
From the obtained fit the *KF* and *AF* factors of the flicker noise model (*NOIMOD-1*) were found as 0.86 and 1.8e-27, respectively.

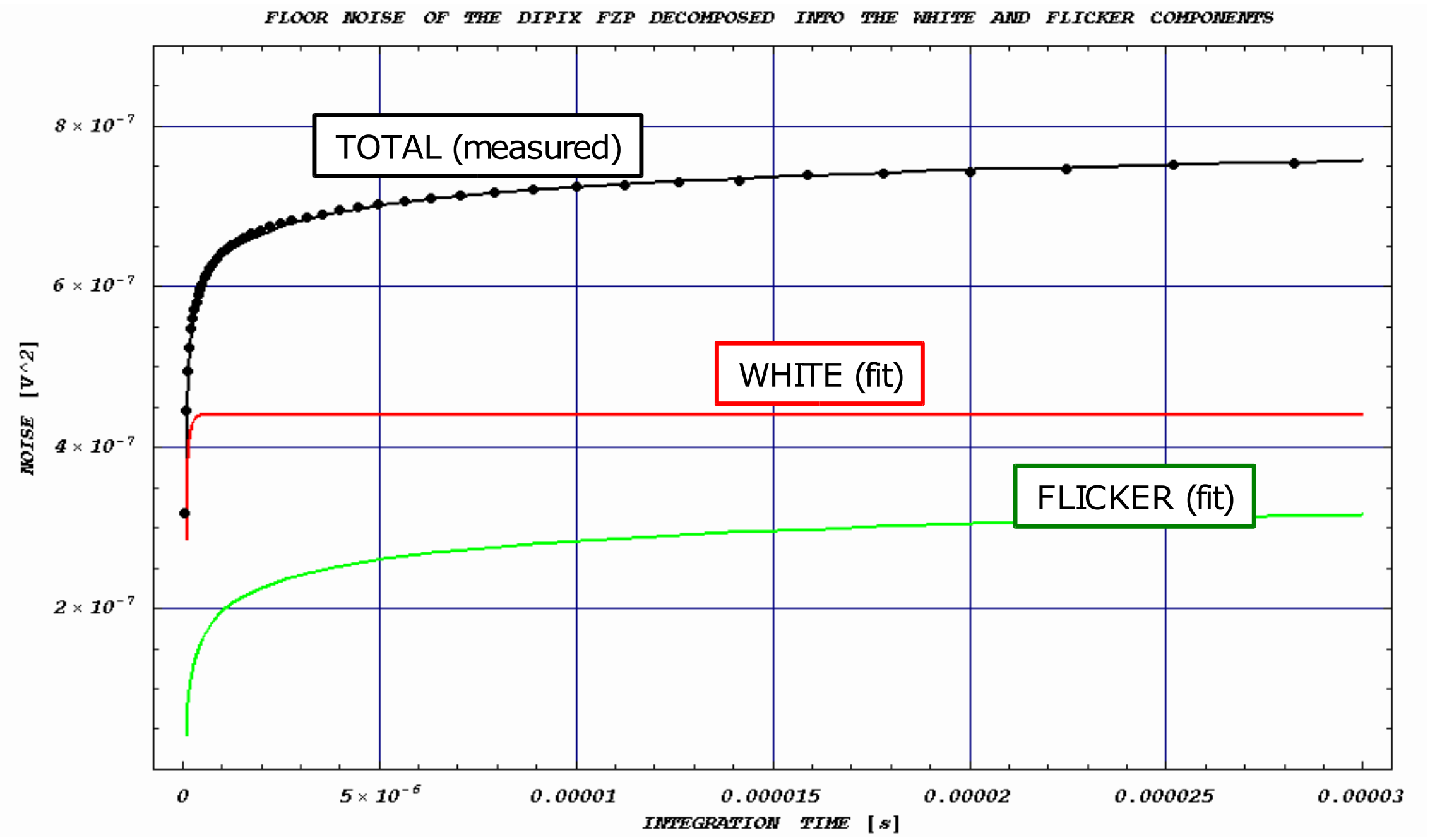

*Drawing 7: The floor noise of the DIPIX FZ-p pixel decomposed into the white and flicker components.*

## 4. Integration time scans

Measurements at different integration times can be used to obtain average values of the pixel leakage current, i.e. unwanted DC current flowing into or out of the pixel.

Here two methods can be considered. In the first one, the slope of the mean pedestal value vs integration time is measured. It is equal: $i_{leak}* (G / c_{pix})$, where G represents voltage gain of the system and the $c_{pix}$ is a pixel capacitance. The $G/c_{pix}$ factor can be found from the Am241 spectra measurements as the *(peak amplitude)/ (corresponding charge)* ratio.

The second method is more complicated. The *ENC* is measured at different integration times what involves both Am241 and noise measurements. Next, the $ENC^2$ vs integration time slope is calculated. The pixel current leak can be found as the slope divided by the electron charge. The "floor noise" must be considered in the measurements, since it contains the *1/f* component , depending on the integration time. The noise measurements are sensitive to external environment (temperature, *EMC* etc), thus the pedestal method is usually preferred.

### 4.1 Measurements of the pix_2012 chip

The *pix_2012*, the first SOI chip designed in Krakow [3], was made on the *CZ-n* wafer. Drawing 8 shows results of integration scan measurements. The pixel leakage current is large (pA range), as expected for the *CZ-n* wafer. Noise and pedestal measurements are consistent.

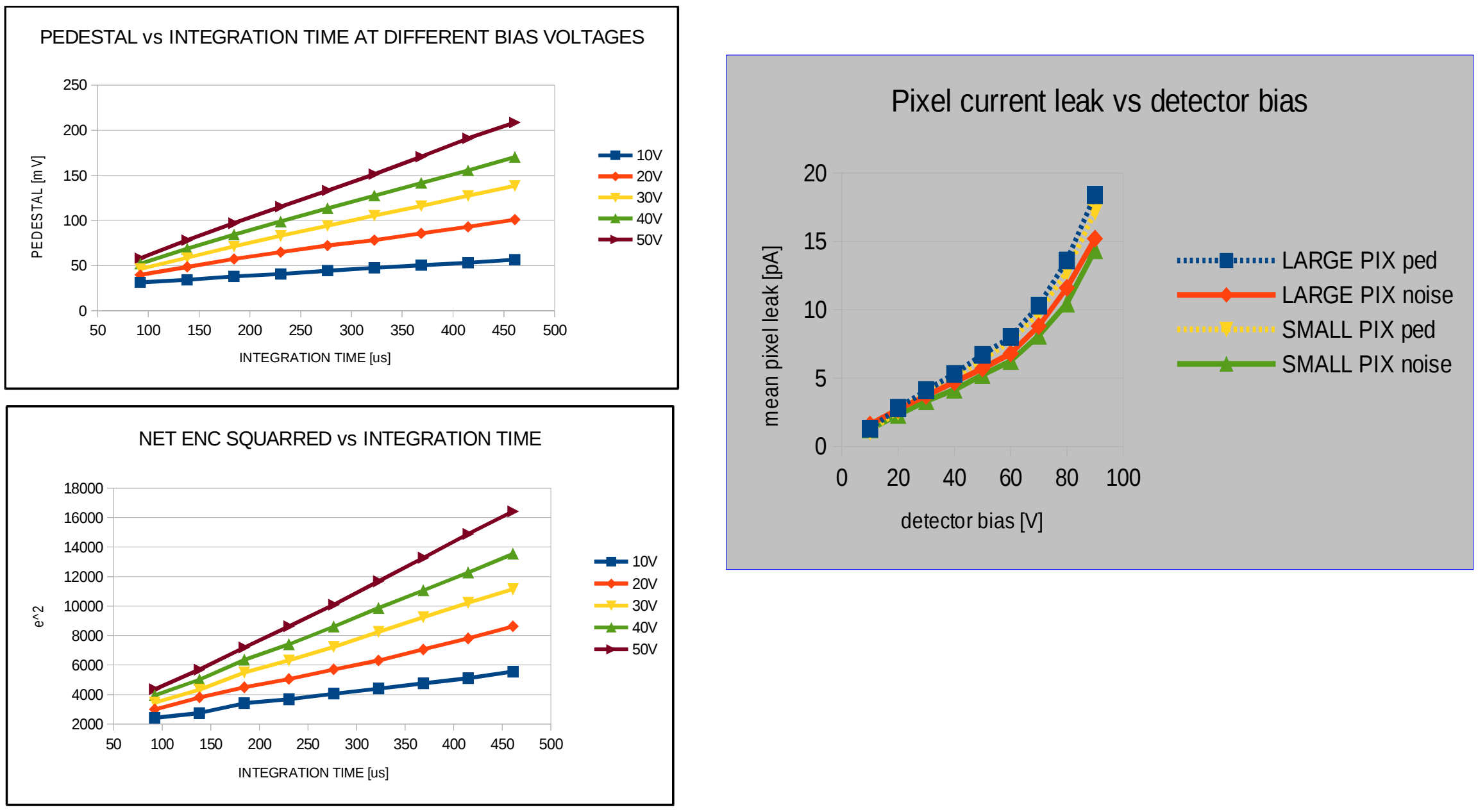

*Drawing 8: Pedestal and squared noise vs integration time. Measurement results of the pix_2012 chip.*

### 4.2 Measurements of the pixel leakage current in the *DIPIX* chips.

Contrary to expectations, already measured pixel leakage currents of the *DIPIX FZ-p* were surprisingly large, in the range of few pA [2]. Due to this, measurements were repeated with extended range of integration times up to 4ms, increased logarithmically. Some precautions were taken about measurements repeatability (cooling, preheating etc). Result, shown in Drawing 9, was a bit surprising. Up to ~700us, a „rapid" grow of pedestals can be observed, i.e. pixel is discharged rapidly. Next the situation stabilizes and the pedestal slopes become linear. Assuming that the true leakage current discharging effect is seen at larger times and using only the data from linear regions, the calculated currents were in the range of hundred fA. Previously measured values were found using data from the region of rapid discharge.

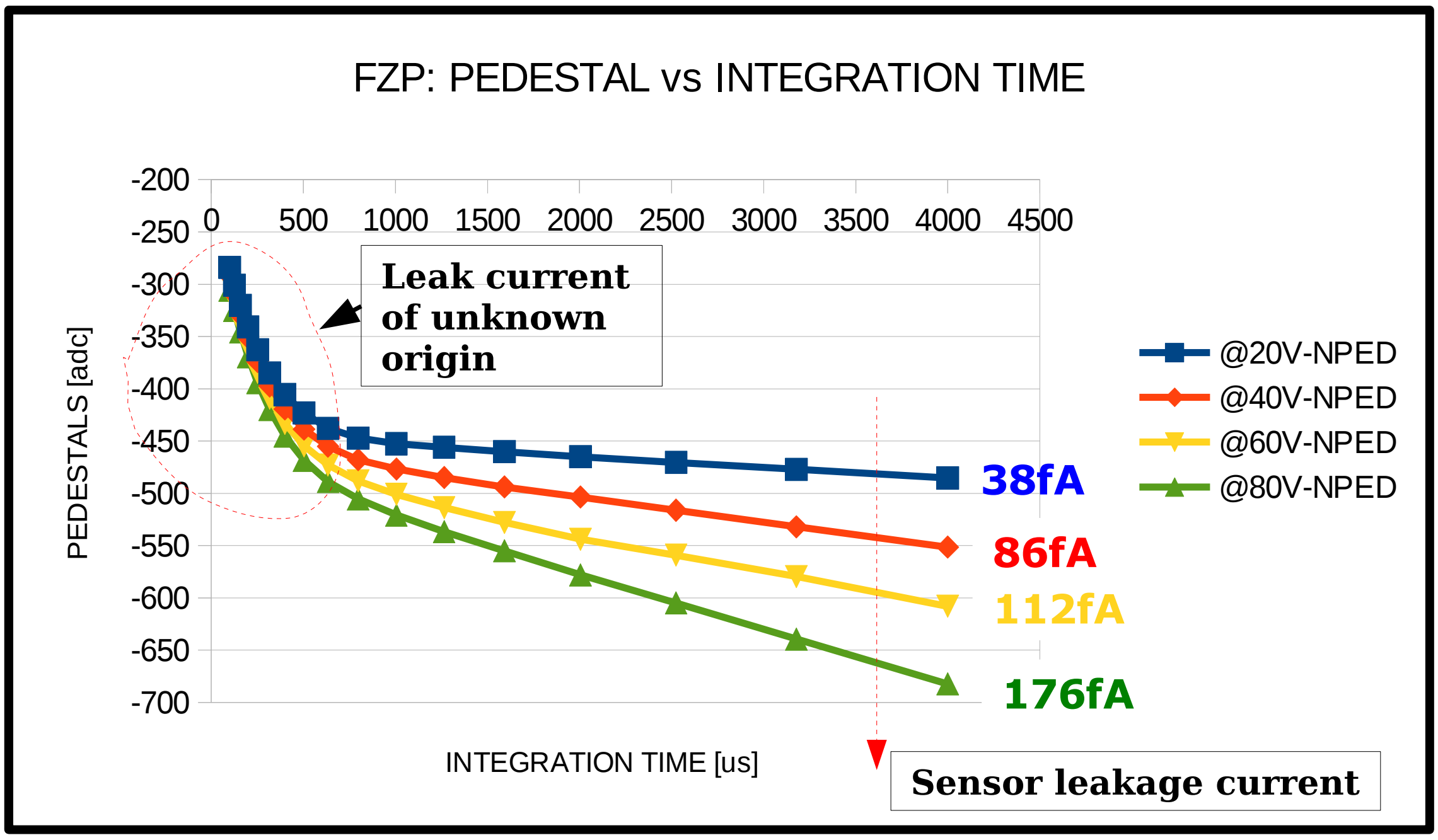

*Drawing 9: Pedestal integration scan of the DIPIX-FZp in the extended range of integration times.*

The extended integration scan procedure was also applied to the *FZn* chip, see Drawing 10. As in the *FZp* case, the similar period of initial pixel discharge is also clearly visible. Leakage currents obtained with data from linear region are smaller then for *FZp*.

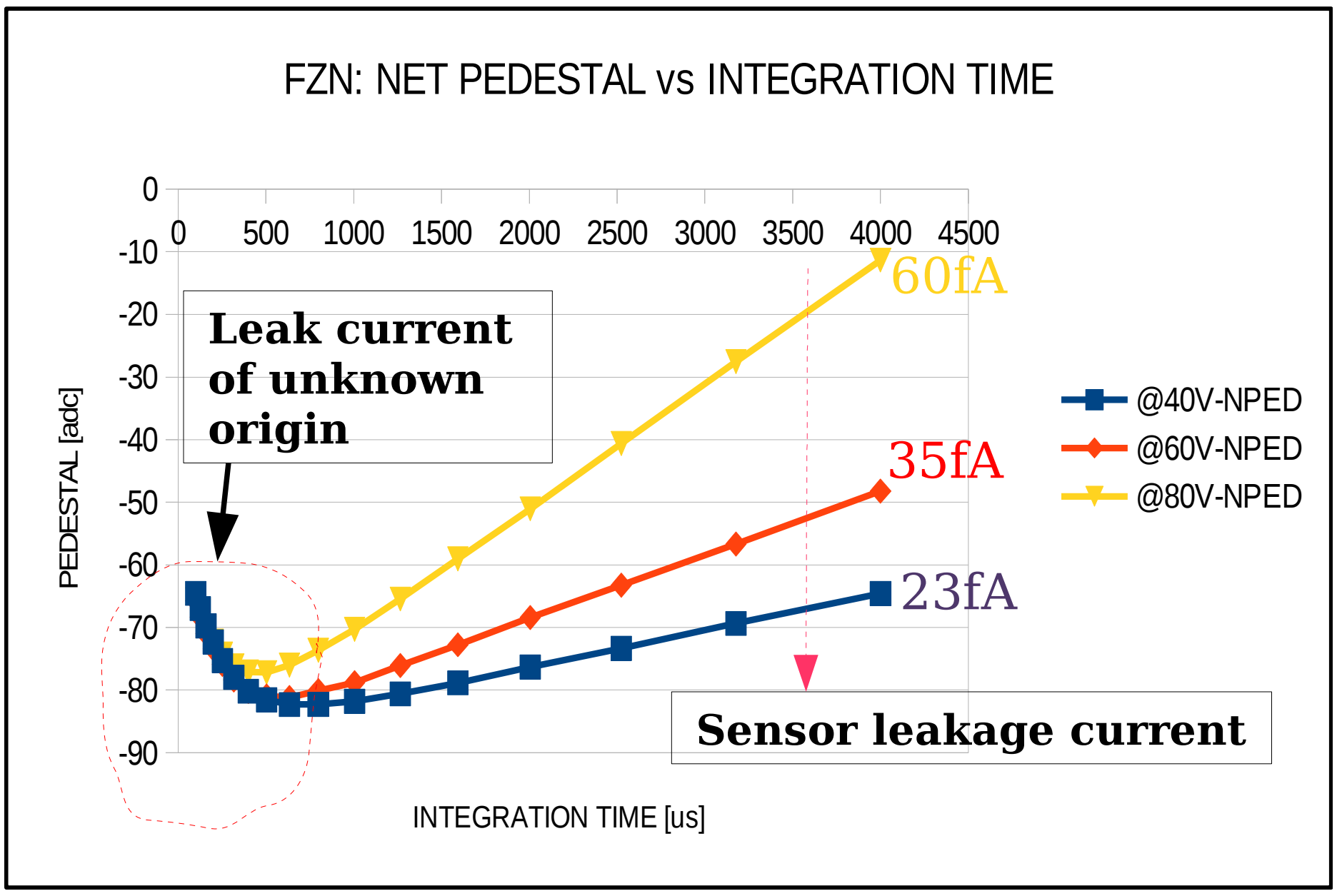

*Drawing 10: Pedestal integration scan of the DIPIX-FZn in the extended range of integration times.*

Finally, the extended pedestal scan was applied to the *CZn* chip. In this case the “rapid discharge” region is barely, but nevertheless visible, hidden in the large leakage current, see Drawing 11.

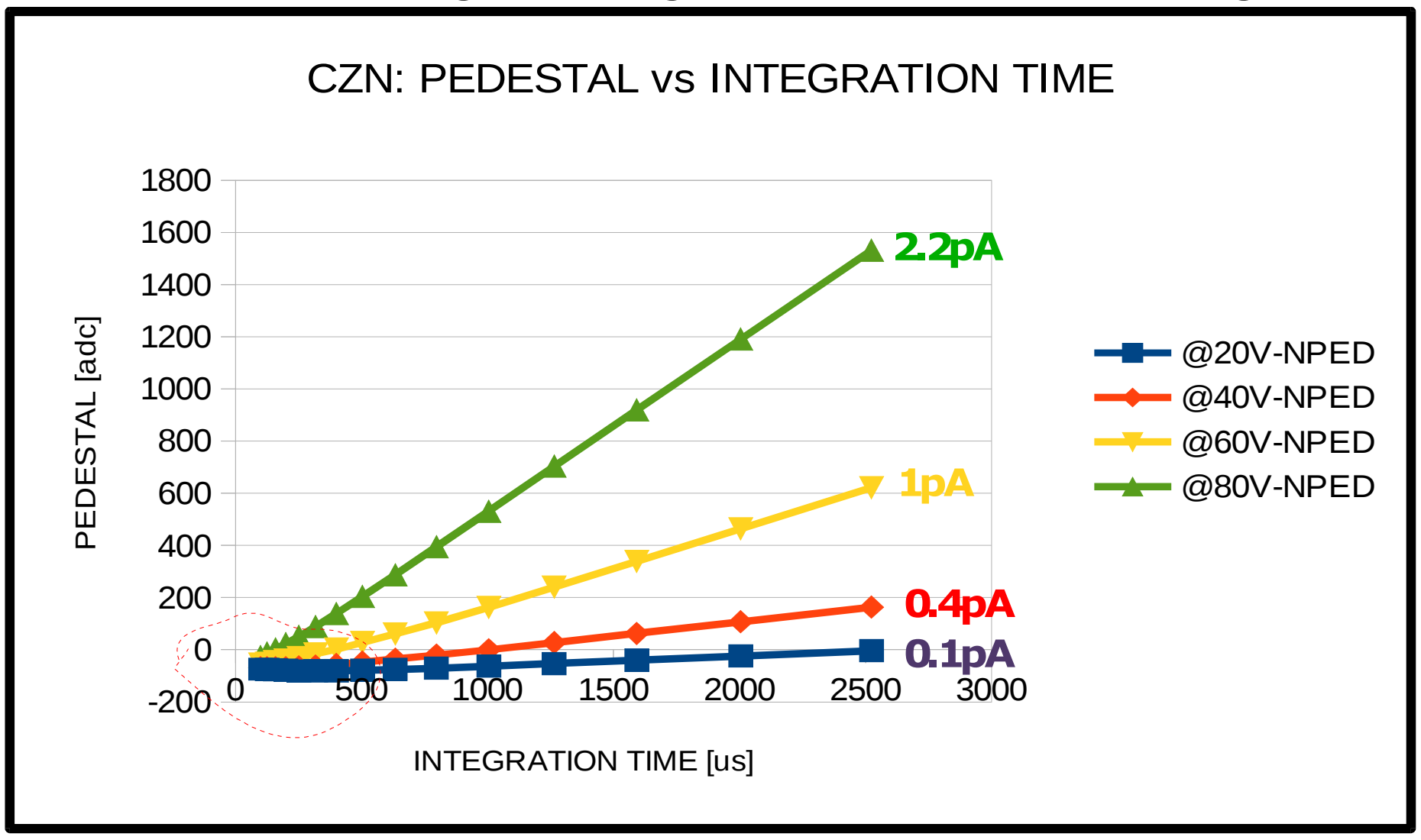

*Drawing 11: Pedestal integration scan of the DIPIX-CZn in the extended range of integration times.*

Using data only from linear regions, a following summary table was made, (values in fA):

**DIPIX PIXEL LEAKAGE CURRENT @80V; ONLY LINEAR REGION**

| CZn | FZn | FZp |
|---|---|---|
| 2200 | 60 | 176 |

### 4.3 Question about the initial pedestal drop origin

In all DIPIX chip versions : *CZn, FZn, FZp* a period of time is observed, when the pedestal is lowering rapidly, what suggests a current flow out of the pixel.
It seems reasonable to assert this effect to the reset switch transistor, made as the *Floating-Body* device. However in this case one would expect an opposite behavior between the *DIPIX-P* and *DIPIX-N* cases.
It was decided, that in the following design submission part o the pixel reset switches will be implemented with the *Body-Tied* transistors. This may help in understanding observed effect.

**Acknowledgments**
The work was supported by Polish National Science Centre (NCN), contract: 2012/07/B/ST2/03752

**References**

[1] P.Seller, *Noise analysis in linear electronic circuits*, Nucl. Instr. And Meth. In Phys. Res. A 376 (1996) 229-241.

[2] M.I. Ahmed et al, *Characterization of high resolution CMOS monolithic active pixel detector in SOI technology* , JINST 10 P05010, 2015

[3] M.I. Ahmed et al, *Prototype pixel detector in the SOI technology*, JINST 9 C02010, February 2014